\documentclass[12pt,twoside]{article}
\usepackage{fleqn,espcrc1,epsfig}

\usepackage{graphicx}
\usepackage[figuresright]{rotating}

\def\Journal#1#2#3#4{{#1} {#2} (#4) #3}

\def\NPA{{Nucl. Phys.} A}
\def\NPB{{Nucl. Phys.} B}
\def\PLB{{Phys. Lett.}  B}
\def\PRL{Phys. Rev. Lett.}
\def\PRD{{Phys. Rev.} D}

\def\ZPA{{Z. Phys.} A}
\def\ZPC{{Z. Phys.} C}

\def\alt{\vcenter{\vbox{\hbox{$\buildrel < \over \sim$}}}}
\def\agt{\vcenter{\vbox{\hbox{$\buildrel > \over \sim$}}}}

\title{\vspace{-1.3cm}
       Hard exclusive photoproduction of $\Phi$ and $J/\Psi$ mesons
\footnote{Talk given by W. Schweiger at the \lq\lq
XVII. European Conference on Few-Body Problems in Physics\rq\rq, Evora, Portugal,
September 2000}
}

\author{
        C.F. Berger
       \address{C.N. Yang Institute for Theoretical Physics, SUNY at Stony
        Brook,\\ Stony Brook, NY 11794-3840, USA},
       B. J\"ager and W. Schweiger
       \address{Institute of Theoretical Physics, University of Graz,\\ 
        Universit\"atsplatz 5, A-8010 Graz, Austria}}
       
\begin{document}

\maketitle

\begin{abstract}
We present predictions for differential cross sections for the reaction 
$\gamma \, p \, \rightarrow  \, \Phi \,  p$ and give an outlook to which extent our
calculations may be generalized to the photoproduction of $J/\Psi$ mesons. Our results are
obtained within perturbative QCD  treating the proton as a quark-diquark system. 
\end{abstract}

\section{INTRODUCTION AND PRESENTATION OF THE MODEL}
The new data from the 93-031 experiment at CEBAF~\cite{An00} have provided precise information
on exclusive photoproduction of $\Phi$ mesons in the momentum-transfer range $1\, \alt\, $ $
\vert t
\vert\, \alt\, 4$~GeV$^2$. It is now interesting to see, whether the momentum transfers
reached in this experiment are already large enough so that the perturbative production mechanism (quark and gluon
exchange) starts to compete with the  non-perturbative one (vector-meson
dominance). From the pertur\-bation-theoretical viewpoint the photoproduction of
$\Phi$s is certainly cleaner and simpler than photoproduction of other vector mesons, like $\rho$s and $\omega$s.
The valence Fock state of the $\Phi$, i.e. the $s$-$\bar{s}$ state, can only be produced via
two-gluon exchange. Quark exchange is OZI suppressed if the strangeness content of the nucleon
is small. Both quark and gluon exchanges are possible, however, in the case of the $\rho$ and
the $\omega$. In the present contribution we are interested in describing $\Phi$ production in
the perturbative regime. We employ a modified version of the hard-scattering
approach~\cite{BL89} in which the proton is considered as a quark-diquark system. This kind
of perturbative model has already been applied successfully to a number of exclusive
photon-induced reactions~\cite{Ja93}$ - $\cite{KSPS97}. A consistent description of baryon
electromagnetic form factors, Compton scattering off baryons, etc. has been achieved in the
sense that the corresponding large momentum-transfer data ($p_\perp^2\, \agt\, 3$~GeV$^2$) are
reproduced with the same set of model parameters. In addition to $\Phi$ production we
investigate whether the same model could be applied to
$\gamma\, p\, \rightarrow \, J/\Psi\, p$ in the few-GeV momentum-transfer region, where
experimental measurements seem to be feasible in the future. 

Within the hard-scattering approach a helicity amplitude $M$ for
the reaction $\gamma \, p\, \rightarrow\, M \, p$ is (to
leading order in $1/p_\perp$) given by the convolution integral~\cite{BL89}
\begin{equation}
M(\hat{s},\hat{t}) \! = \! \int_0^1 \! \! dx_1 dy_1 dz_1
{\phi^M}^{\dagger}(z_1)
{\phi^p}^{\dagger}(y_1)
\widehat{T}(x_1,y_1,z_1;\hat{s},\hat{t})
\phi^p(x_1) \, . \label{convol}
\end{equation}
The distribution amplitudes $\phi^H$ are probability amplitudes for finding the valence Fock
state in the hadron $H$ with the constituents carrying certain fractions $x_i, y_i, z_i, \,
i=1,2$, of the momentum of their parent hadron. In our model the valence Fock state of an
ordinary baryon is assumed to consist of a quark ($q$) and a diquark ($D$). The hard
scattering amplitude $\widehat{T}$ is calculated perturbatively in collinear
approximation and consists in our particular case of all possible tree diagrams contributing
to the elementary scattering process  $\gamma\,q\, D\, \rightarrow\, Q \, \bar{Q}\, q \, D$.
The Mandelstam variables $s$ and $t$ are written with a hat to indicate that they are defined
for vanishing hadron masses. Helicity labels are neglected in Eq.~(\ref{convol}). Our calculation of the
hard-scattering amplitude involves an expansion in powers of $(m_H/\sqrt{\hat{s}})$ which is
performed at fixed scattering angle. We keep only the leading order and next-to-leading order
terms in this expansion. Hadron masses, however, are fully taken into account in flux and
phase-space factors. For a more detailed explanation of our treatment of mass-effects which,
by the way, also guarantees the gauge invariance of the mass-correction terms we refer to
Ref.~\cite{BS00}.

The model, as applied in Refs.~\cite{Ja93,Kro93,Kro96,KSPS97}, comprises scalar ($S$) as well
as axial-vector ($V$) diquarks. For the Feynman rules of electromagnetically and strongly
interacting diquarks, as well as for the choice of the quark-diquark distribution amplitudes of
octet baryons we refer to Ref.~\cite{Ja93}. Here it is only important to mention that the
composite nature of diquarks is taken into account by diquark form factors. For further details
of the diquark model we refer to the publication of R.~Jakob et al.~\cite{Ja93}. The numerical
values of the model parameters for the present study are also taken from this paper where they were fixed by fitting elastic electron-nucleon scattering data. The new
ingredient, namely the $\Phi$-meson distribution amplitude, is adopted from a paper of
Benayoun and Chernyak~\cite{BC90}. It fulfills QCD sum-rule constraints. In order to treat
$\Phi$ and $J/\Psi$ production on the same footing we adapt this distribution
amplitude~\cite{BS00} by attaching an exponential factor which provides a flavor dependence in
accordance with heavy-quark effective theory. 

\vspace{-0.3cm}

\section{RESULTS}
\begin{figure}[t!]
\begin{center}
\epsfig{file=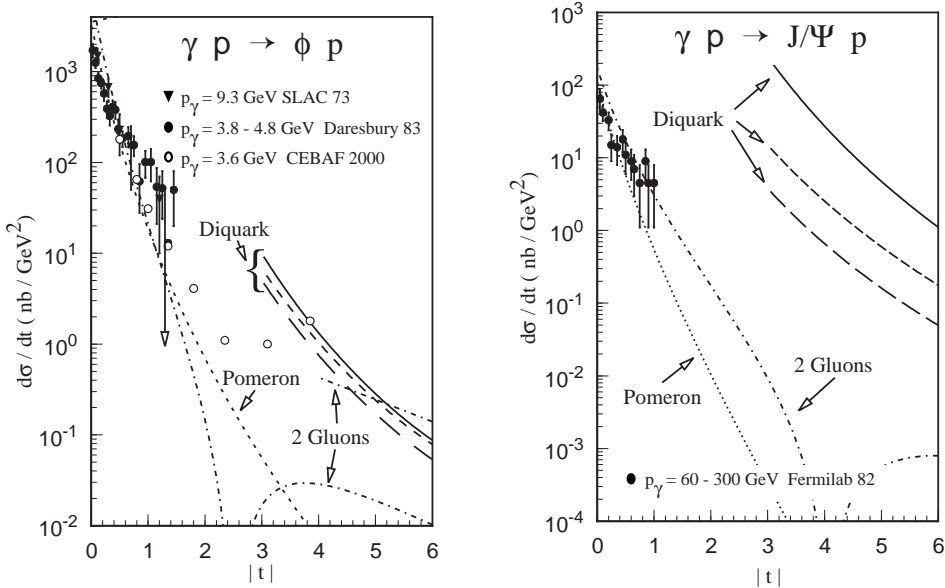,height=8.0cm,clip=}
\end{center}
\vspace{-1.0cm}
\caption{Differential cross section for the photoproduction of the
$\Phi$-$p$ ({\em left figure}) and the $J/\Psi$-$p$ ({\em right figure}) final
states for photon lab-energies of 6 and 150 GeV, respectively. Diquark model predictions
with different treatment of hadron-mass effects correspond to the long-dashed (masses
neglected), the solid (first-order mass corrections included), and the short-dashed
(first-order mass corrections + meson mass in non-singular propagators included) lines. Cross
sections resulting from the Pomeron-exchange mechanism (dotted line)
and a two-gluon-exchange model~{\protect \cite{La95}} (dash-dotted line) are plotted for
comparison. $\Phi$-data are taken from Ballam et al.~{\protect \cite{Bal73}}, Barber et
al.~{\protect \cite{Ba83}}, and Anciant et al.~{\protect \cite{An00}}. The $J/\Psi $ data
points have been obtained by Binkley et al.~{\protect \cite{Bi82}} by averaging the measured
cross sections over photon lab-energies between 60 and 300 GeV.}
\label{result}
\vspace{-0.7cm}
\end{figure}

Fig.~\ref{result} shows (unpolarized) differential cross sections for the photoproduction
of $\Phi$ (left) and $J/\Psi$ (right) mesons. In the absence of data at large
(transverse) momentum transfer, i.e. $t$ {\bf and} $u$ large, we are not able to directly
confront the diquark-model predictions (solid line) with experiment. However, comparing with
the old experimental data points which are available for photon lab energies between 3.8 and
9.3 GeV, but only small momentum transfers ($t\, \alt\, 1.5$~GeV$^2$), one can recognize that
our results seem to provide a reasonable extrapolation in case of the $\Phi$, but fail
completely for the $J/\Psi$. For comparison we have also plotted the $\Phi$ photoproduction
data from the CEBAF 93-031 experiment which have been published most recently~\cite{An00}. The
$t$-dependence of our model is clearly at variance with these data, indicating that the
energies and momentum transfers reached in the CEBAF 93-031 experiment are still too low to
explore the perturbative regime. This is not surprising since the maximum transverse momentum
transfer reached in this experiment corresponds to $t$ and $u$ values of about 2.3~GeV$^2$. 
As can be seen from Fig.~\ref{result} the forward cross section for $\Phi$ production is
reasonably well reproduced by simple Pomeron  phenomenology (dotted line). A QCD-inspired 
version of the Pomeron exchange, in which the Pomeron is replaced by two non-perturbative 
(abelian) gluons, has been proposed by Laget and Mendez-Galain~\cite{La95} to link the 
diffractive with the hard-scattering domain (dash-dotted line). If the two gluons are only
allowed to couple to the same quark in the proton,  the two-gluon-exchange cross section 
exhibits a characteristic node around $t \approx 2.5$~GeV$^2$ which can be understood as an
interference effect between the two Feynman  diagrams which enter the photoproduction
amplitude. This node is completely washed out if the two  gluons are also allowed to couple to
different quarks in the proton~\cite{La00} (upper dash-dotted line). Actually, for $\vert t
\vert \agt 4$~GeV$^2$ the result of Laget becomes  comparable with the diquark-model result.
This is not surprising since we know from the perturbative analysis that  the hard-scattering
mechanism, i.e. contributions from diagrams without loops in which all the  hadronic
constituents are connected via gluon exchange, should become dominant for large values of $t$
and $u$. Also the two-gluon exchange model is not able to account for the flattening of the
cross section data around $\theta_{\hbox{cm}}=90^{\circ}$ and the rise in backward direction.
A non-perturbative mechanism, namely $u$-channel nucleon exchange, has to be introduced to
explain this feature~\cite{An00,La00}. 

The problem that $\Phi$ production is well reproduced, whereas $J/\Psi$ production is
overestimated does not only occur within our model for the hard-scattering region, but is
also encountered within the pomeron-exchange model and the two-gluon-exchange model. For both
models the situation in $J/\Psi$ production is improved by arguing that the coupling of the
pomeron to a charmed quark is weaker than the coupling to a strange quark. The failure of our
model for $J/\Psi$ production can be traced back to the \lq\lq hadron-mass corrections\rq\rq\
which enter the cross section only via the hadron-helicity-flip amplitudes (apart from flux
and phase-space factors). Whereas the two hadron-helicity conserving amplitudes provide
already about 60\%  of the cross section in case of the $\Phi$, they are negligible in case of
the $J/\Psi$ (solid line vs. long-dashed line). By comparing the kinematics of the two plots in
Fig.~\ref{result} one observes that the relevant parameter of our mass expansion, namely 
$(m_M/\sqrt{\hat{s}})$, is in both cases small, but both reactions are considered within very
different angular ranges. $\vert t \vert$ values between 3 and 6~GeV$^2$ correspond to a
scattering angle between 70 and 110 degrees for a photon lab-energy of 6~GeV, but to
nearly-forward scattering (between 10 and 20 degrees) for a photon lab-energy of 150~GeV.
Keeping in mind that our mass expansion is an expansion at {\bf fixed} angle, the different
magnitude of mass corrections in $\Phi$ and $J/\Psi$ production becomes obvious. The expansion
coefficient of the first mass-correction term is angular dependent and large for small
scattering angles. This indicates that perturbation theory for $J/\Psi$ production is only
trustworthy at large enough energies and scattering angles as long as the $J/\Psi$ mass is not
fully taken into account. A possible way to apply the perturbative approach to $J/\Psi$
production at a few GeV of momentum transfer would be  to refrain from the expansion
in $(m_{J/\Psi}/\sqrt{\hat{s}})$. The full inclusion of the  $J/\Psi$ mass, however, entails
complications with the numerical treatment of propagator singularities in the convolution 
integral, Eq.~(\ref{convol}), and with the analytical expressions for the amplitudes which may become
very lengthy. Nevertheless, we are presently attempting to go into this direction. As a first
step we have tried to include the meson mass in the denominator of the non-singular  heavy
quark propagators. As can be observed from Fig.~\ref{result} (short-dashed line) this indeed
leads to a reduction and a slight flattening of the cross section. These effects are, of
course, more pronounced for the $J/\Psi$ than for the $\Phi$, but are welcome in both cases.
It remains to be seen whether the full inclusion of the meson mass in the denominators of the
singular propagators and also in the numerator of the Feynman amplitudes will finally provide
results which extrapolate the small-$t$ HERA data for the photoproduction of $J/\Psi$ in a
reasonable way.

\end{document}